\documentstyle[12pt]{article}

\def\bseq{\begin{subequation}} 
\def\eseq{\end{subequation}}
\def\bsea{\begin{subeqnarray}} 
\def\esea{\end{subeqnarray}}

\def\beq{\begin{equation}}
\def\eeq{\end{equation}}
\def\eea{\end{eqnarray}}
\def\bq{\begin{quote}}
\def\eq{\end{quote}}

\newcommand{\EQ}{\begin{equation}}
\newcommand{\EN}{\end{equation}}
\newcommand{\bea}[1]{\begin{eqnarray}\label{#1}}
\newcommand{\ena}{\end{eqnarray}}

\renewcommand{\a}{\alpha}
\newcommand{\ad}{{\dot\alpha}}
\renewcommand{\b}{\beta}
\newcommand{\bd}{{\dot\beta}}

\renewcommand{\d}{\delta}
\newcommand{\th}{\theta}

\renewcommand{\L}{\Lambda}

\newcommand{\f}{{\phi}}
\newcommand{\fb}{\bar {\phi}}
\newcommand{\phibar}{\bar{\phi}}

\newcommand{\FF}{{\cal F}}
\newcommand{\HH}{{\cal H}}
\newcommand{\Wbar}{\bar {W}}

\newcommand{\del}{\partial}
\newcommand{\Del}{\nabla}
\newcommand{\Delb}{{\bar\nabla}}

\newcommand{\shalf}{\hbox{$\frac12$}} 
\newcommand{\squart}{\hbox{$\frac14$}} 

\def\bop#1{\setbox0=\hbox{$#1M$}\mkern1.5mu
 \vbox{\hrule height0pt depth.04\ht0
 \hbox{\vrule width.04\ht0 height.9\ht0 \kern.9\ht0
 \vrule width.04\ht0}\hrule height.04\ht0}\mkern1.5mu}
\def\Box{{\mathpalette\bop{}}}

\def\eed{{\hbox{C\kern-.3em I}}}

\begin{document}
\begin{titlepage}
\begin{flushright}
ITP-SB-95-55\\
BRX-TH-387\\
THU-95/32\\
hep-th/9601115\\
January 1996
\end{flushright}

\begin{center}
{\large\bf NONHOLOMORPHIC CORRECTIONS TO THE ONE-LOOP $N=2$ SUPER
YANG-MILLS ACTION}\\
\vspace{1.0cm}

\normalsize
B. de Wit\footnote{email:bdewit@fys.ruu.nl}\\{\it Institute for
Theoretical Physics}\\
{\it Utrecht University}\\{\it Princetonplein 5, 3508 TA,
Utrecht, The Netherlands}\\
\bigskip

M.T. Grisaru\footnote{email:grisaru@binah.cc.brandeis.edu}\\{\it
Physics Department}\\{\it Brandeis University}\\
{\it Waltham, MA 02254, USA}\\
\bigskip

M.\ Ro\v cek\footnote{email:rocek@insti.physics.sunysb.edu}\\
{\it Institute for Theoretical Physics\\ State University of New York\\
 Stony Brook, NY 11794-3840, USA\\}
\end{center}
\vspace{1.0cm}
\begin{abstract}
\bigskip
In addition to the familiar contribution from a holomorphic function
$\FF$, the K\"ahler potential of the scalars in the nonabelian $N=2$
vector multiplet receives contributions from a real function $\HH$. We
determine the latter at the one-loop level, taking into account
both supersymmetric matter and gauge loops. The function $\HH$
characterizes the four-point coupling of the $N=2$ vector-multiplet
vectors for constant values of their
scalar superpartners. We discuss the consequences of our results.
\end{abstract}
\end{titlepage}

\newpage

Low-energy effective actions reveal much of the interesting
information about a theory; for string theory and supersymmetric gauge
theories, they encode non-perturbative information about the mass
spectrum and the static couplings. Recently, considerable attention
has been given to the study of effective actions for $N=2$
supersymmetric Yang-Mills theories. This has followed Seiberg and
Witten's \cite{sw} construction of the {\em exact} low-energy
effective action for the $N=2$ $SU(2)$ theory (which breaks down to an
abelian phase). The work by Seiberg and Witten and others is based on
the description of the $N=2$ superspace effective action by means of a
chiral integral of a holomorphic function $\FF(W)$, where $W$ is the
$N=2$ gauge superfield strength of the unbroken $U(1)$  \cite{f}.  In terms of
$N=1$ superfields, this has the form\beq
S=\frac{1}{16\pi} {\cal I}m \left[ \int d^4x \,d^2\th\,\,
\FF_{\f \f}(\f )\, (\shalf W^\a W_\a)
+\int d^4x \,d^2\th\, d^2{\bar\th}\,\, \fb \FF_{\f} (\f)\right]\ .
\label{S0}
\eeq
Here $\f$ denotes the chiral superfield\footnote{Except when discussing
Feynman rules, we work with covariantly chiral and antichiral
superfields to avoid writing explicit factors of $e^V$.} that is the
lowest $N=1$ superspace component of the $N=2$ (unbroken)
$U(1)$ gauge multiplet, and $W_\a$ is the $N=1$ gauge superfield
strength. This effective action is consistent with (the rigid version
of) special geometry \cite{f}. In particular, the K\"ahler potential
-- the chiral superfield Lagrangian -- is given by
\beq
K(\f ,\fb)=\frac{1}{32 \pi i}(\fb \,{\FF}_{\f}-\f \,\bar{\FF}_{\fb}) \ .
\label{Kf}
\eeq
Classically,
\beq
\FF (\f ) = \frac{4\pi i}{g^2}\, \frac{\f^2}2\ .
\label{class}
\eeq

In this paper we examine the form of the {\em non}abelian low-energy
effective action. The kinetic terms for the component fields of a
chiral superfield $\f$ come from an $N=1$ superspace Lagrangian, the
K\"ahler potential $K$. Since $K$ is a function of $\f$ and $\fb$ that
does not depend on their (spinor) derivatives, we can calculate its
loop corrections in much the same way as one calculates the effective
potential in component theories. We compute one-loop contributions to
$K(\f,\fb)$ induced by both $N=2$ vector multiplets and
hypermultiplets. Somewhat unexpectedly, in the nonabelian case the
K\"ahler potential cannot be written in the form $K(\f ,\fb )\propto
{\rm Im} ~[\fb \FF_{\f}]$ ({\it cf.} (\ref{Kf})), so that it is not
determined solely by a holomorphic function $\FF(\f)$. The additional
terms originate from a real function $\HH (W,\bar{W})$ of the $N=2$
Yang-Mills superfield strength $W$, which is integrated with the full
$N=2$ superspace measure.\footnote{Such a function has recently been
discussed by Henningson \cite{h} in the abelian limit, where it does
not contribute to the effect that we study.} We deduce the one-loop
contribution to $\HH$ from our computation of $K(\f,\fb)$.

We begin with the classical action for $N=2$ supersymmetric systems
written in $N=1$ superspace:
\bea{a}
S &=& \frac{1}{4g^2}\left[ \int d^4x \,d^2 \th\, {\rm Tr}
(\shalf W^\a W_\a )+\int d^4x \,d^4 \th\,{\rm Tr}( \fb \,e^{-V}
\f \, e^V ) \right] \nonumber \\
&&~~~~~~ +\int d^4x \,d^4 \th\, (\bar{Q}\, e^V Q +\tilde{Q} \, e^{-V}
\bar{\tilde Q}) + \Big(i\int d^4x\,d^2 \th\,\tilde Q \phi Q +
h.c.\Big) \,,
\eea
where $\f$ and $\fb$ are Lie-algebra valued,
$\f= \f^A T_A$, $\fb= \fb^A T_A$, with $[T_A,T_B]= i f_{AB}^C T_C$, while
$Q$ and $\tilde Q$ are in mutually conjugate representations $R$ and
$\tilde R$. The superfields $W_\a$ and $\f$ are $N=1$ superfield
components of the reduced chiral $N=2$ superfield $W$ which describes
the $N=2$ vector multiplet (see below for details), and
$W_\a\equiv i\bar D^2(e^{-V}D_\a e^V)$ depends on $V$ in the usual
manner. The chiral $N=1$ superfields $Q$ and $\tilde Q$ together
describe $N=2$ hypermultiplets. We use the conventions of {\em
Superspace} \cite{Superspace}.

\let\picnaturalsize=N
\def\picsize{4.0in}
\def\picfilename{nonholo.eps}
\ifx\nopictures Y\else{\ifx\epsfloaded Y\else\input epsf \fi
\let\epsfloaded=Y
\centerline{\ifx\picnaturalsize
N\epsfxsize \picsize\fi \epsfbox{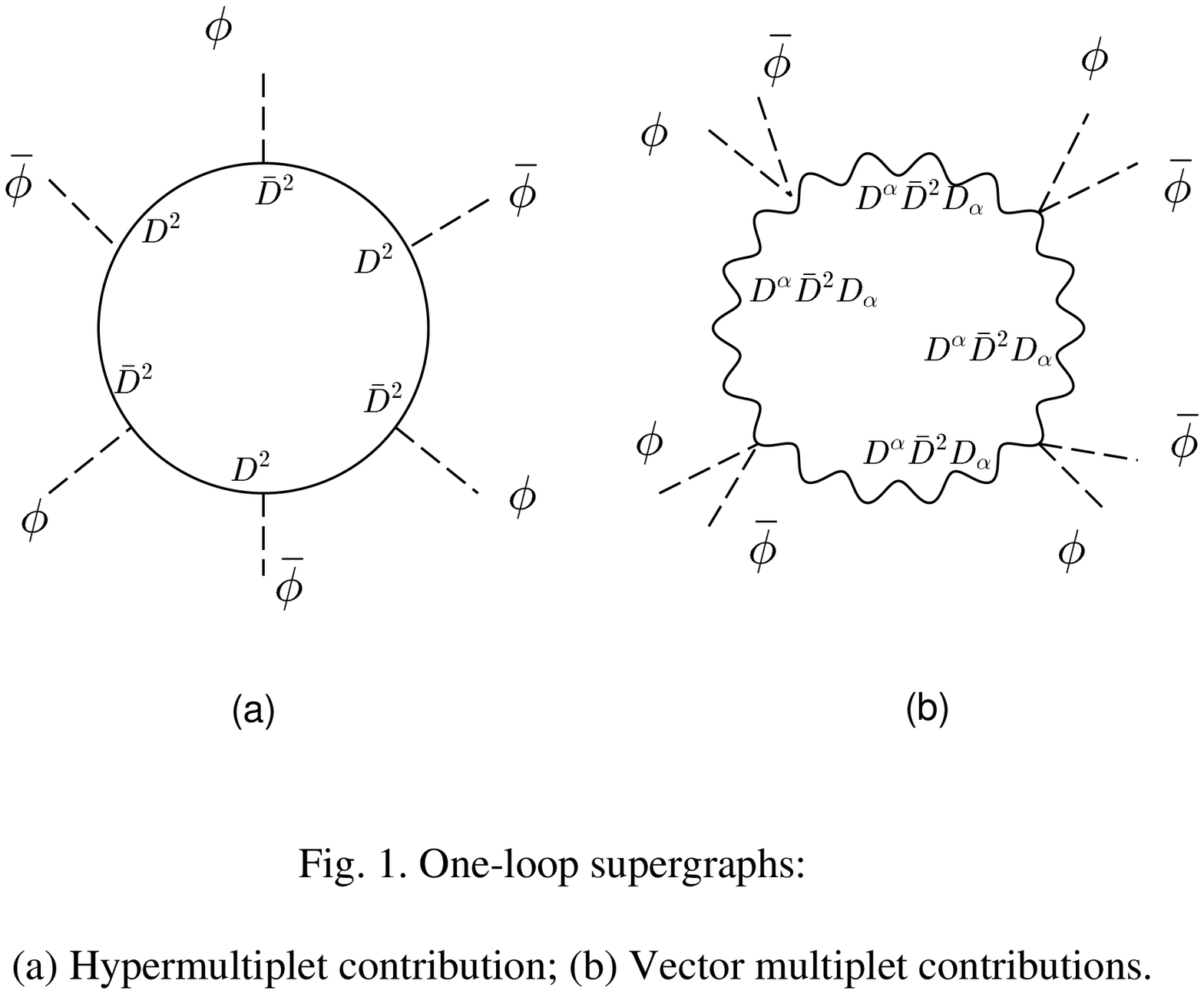}}}\fi

At the one-loop level we consider all diagrams with external
$\f$, $\fb$ lines, but when doing $D$-algebra, drop terms with spinor
or space-time derivatives on the external lines. The contributions to
the K\"ahler potential for the chiral superfields $\f$ come from
internal lines corresponding to the $(Q,\tilde Q)$ hypermultiplets and
the $N=2$ vector multiplet itself. The chiral multiplets $Q$ and
$\tilde Q$ give rise to (with a $d^2\th\, d^2\bar{\th}$ integration)
\beq\label{QQ}
{\rm Tr}_R \int \frac{d^4p}{(2\pi )^4} \frac{-1}{p^2}
\sum_{n=1}^{\infty} \frac{1}{n}
\left( \frac{\phibar \phi}{-p^2}\right)^n = \frac{1}{(4\pi )^2}
{\rm Tr}_R
\int_0^{\infty}
dp^2 \ln \left(1+\frac{\phibar \phi}{p^2} \right)\, ,
\eeq
where we summed over diagrams with $n$ $\f$ and $n$ $\fb$ external
lines, $2n$ propagators $-1/\Box$, and $n$ factors of $D^2$,
$\bar{D}^2$ (from the chiral superfield vertices), $n-1$ of which
cancel propagators in the course of $D$-algebra. Integrals are
evaluated in Euclidean space. The momentum integral is divergent and
needs to be regularized; all cut-off dependence contributes only to
renormalizations of the original action (\ref{a}), and the interesting
terms come only from the lower limit of the integral:
\beq\label{hyp}
K_{Q}(\f,\fb) =
-\frac{1}{(4\pi )^2} {\rm Tr}_R\;\Big[ \fb \f \ln {\fb
\f\over \L^2} \Big] \ .
\eeq
where $\L$ is a renormalization scale.

For the $N=2$ vector multiplet itself, in a general gauge with a
supersymmetric gauge-fixing term $\a^{-1}(D^2V)(\bar D^2V)$, both $V$
and $\f$ contribute to the K\"ahler potential.\footnote{For the $N=1$
supersymmetric gauge fixing term that we use, there is no coupling of
the ghosts to the external fields $\f$.}  However, it is simplest
to calculate in supersymmetric Landau gauge ($\a=0$), where only the
$N=1$ vector multiplet $V$ enters in the loop: the $V$-propagator
carries a factor $D^\a\bar {D}^2D_\a$, and factors of $D^2$ or
$\bar{D}^2$ from internal $\f$, $ \fb$ lines annihilate the mixed-loop
contributions stemming from vertices ${\rm Tr}\,(\fb [V, \f])$ (recall
that we drop terms with derivatives on external lines). Thus, only
the vertices ${\rm Tr}\,(\fb [V, [V, \f ]])$ contribute. The result is
independent of $\a$: the mixed-loop contribution cancels the $\a$
dependence of the $V$-propagator in the pure $V$ loops. Summing over
diagrams with $n$ ${\rm Tr}\,( \fb [V, [V, \f ]])$ vertices and
$n$
$V$-propagators $-D^\a\bar {D}^2D_\a/\Box^2$, we obtain
\beq
{\rm Tr}_{(ad)} \int \frac{d^4p}{(2\pi )^4} \frac1{p^2}
\sum_1^{\infty}\frac1{n}
\left( \frac{\f\fb+\fb\f}{-2p^2}\right)^n = \frac{-1}{(4\pi )^2}
{\rm Tr}_{(ad)} \int_0^{\infty}
dp^2 \ln \left(1+\frac{\f\fb+\fb\f}{2p^2} \right)\ .
\eeq
Doing the integral and renormalizing as before, we obtain
\beq\label{vec}
K_{V}(\f,\fb) =
\frac1{(4\pi )^2} {\rm Tr}_{(ad)} \Big[\frac{\f\fb+\fb\f}2 \ln
\frac{\f\fb+\fb\f}{2\L^2}\Big] \ ,
\eeq
We can check that our renormalizations of (\ref{hyp}) and of
(\ref{vec}) are consistent by considering the $N=4$ supersymmetric
case. This is the theory with precisely one
hypermultiplet in the adjoint representation, and we find that
the dependence on the renormalization scale cancels, as it should.

We observe that the results in (\ref{hyp}, \ref{vec}) are in
conflict with special geometry except in the abelian case when $\f$
and $\fb$ commute; in that case, it is easy to recover the following
expression for the holomorphic function $\FF$,
\beq\label{hol}
\FF(\f) = \frac{-i}{2\pi} \left( {\rm Tr}_R \Big[\f^2 \ln
\frac{\f^2}{\L^2}\Big] -{\rm Tr}_{(ad)}
\Big[\f^2 \ln \frac{\f^2}{\L^2}\Big] \right) \ ,
\eeq
where $\L$ has been suitably rescaled.

We explicitly evaluate the K\"ahler potential for $SU(2)$, with
$Q,\tilde{Q}$ in the fundamental and adjoint representations. In the
first case we find
\beq\label{fun}
K_{Q}^{fund} =
\frac{-1}{(8\pi)^2}\,\left(\fb\cdot\f \ln \frac{\f^2 \fb^2}{16\L^4} +
i|\fb\times\f |\ln \frac{
\fb\cdot\f +i|\fb\times \f |}{ \fb\cdot\f -i |\fb\times\f |}\right)\ ,
\eeq
where $|\phibar \times \phi |=\sqrt{\f^2\fb^2-(\fb\cdot\f)^2}$,
$\f^2=\f^A\f^A$, and $\fb\cdot\f=\fb^A\f^A$.
For the adjoint representation we find
\beq\label{ad}
K_{Q}^{adj} = \frac{-1}{(4\pi)^2}\, \left(\fb\cdot\f \ln \frac{\f^2
\fb^2}{\L^4} +2\fb\cdot\f \ln \frac{\fb\cdot\f}{\sqrt{\f^2 \fb^2}
}\right)\ .
\eeq
The contribution from the vector multiplet itself is
\bea{v}
K_{V}&=&\frac1{(4\pi )^2}\left[\fb\cdot\f \ln \frac{\f^2\fb^2}{\L^4}
+\fb\cdot\f \ln \frac{\fb\cdot\f} {\sqrt{\f^2 \fb^2}} \right.
\nonumber\\&&\left. \qquad\qquad\right. \nonumber\\
&&\left.\qquad\qquad +\shalf \left(\fb\cdot\f+\sqrt{\f^2\fb^2}
\right)\ln\frac{\fb\cdot\f+\sqrt{\f^2\fb^2}}{2\sqrt{\f^2\fb^2}}\right.
\nonumber\\&&\left. \qquad\qquad\right. \nonumber\\
&&\left.\qquad\qquad +\shalf\left(\fb\cdot\f-\sqrt{\f^2\fb^2}\right)
\ln\frac{\fb\cdot\f-\sqrt{\f^2\fb^2}}{2\sqrt{\f^2\fb^2}}\,\right]\ .
\eea
In the abelian case, only the first term in (\ref{fun}), (\ref{ad})
and (\ref{v}) survives.

Thus, for the general nonabelian case, the one-loop contributions to the
K\"{a}hler potential do not have the naively anticipated form.
However, by considering higher-dimension local terms in the
$N=2$ superspace effective action, we will show that there is no
contradiction with $N=2$ supersymmetry.

To determine the $N=2$ superspace terms that give rise to the
contributions above, we briefly review $N=2$ super Yang-Mills theory in
superspace. In our notation, the $N=2$ spinor coordinates form a
doublet $\{\th^{a\a}\}=\{\th^{1\a},\th^{2\a}\}$ which transforms under
a rigid $SU(2)$ group unrelated to the gauge group. We also have their
complex conjugates $\{\bar\th^\ad_a\}$, and spinor derivatives
corresponding to all the coordinates. Chiral integrands are evaluated
with the chiral measure $d^4\th = (d^2 \th^1)(d^2\th^2)$; the full
measure is denoted by $d^4\th \,d^4\bar{\th}$. The rigid $SU(2)$ indices
are raised and lowered by the antisymmetric invariant tensor $C_{ab}$,
with $C_{12}= C^{12}=1$. In this context, $N=2$ super Yang-Mills theory
is described by gauge-covariant spinor derivatives satisfying the
constraints
\bea{cons}
\{\Del_{a\a},\Del_{b\b}\}&=&iC_{ab}C_{\a\b}\,\bar W\ ,\nonumber\\
\{\Delb^a_\ad,\Delb^b_\bd\}&=&iC^{ab}C_{\ad\bd}\,W\ ,\nonumber\\
\{\Del_{a\a},\Delb^b_\bd\}&=&i\d_a^b\, \Del_{\a\bd}\ ,
\eea
where $W,\bar W$ are chiral (antichiral) scalar superfield strengths,
respectively. The Bianchi identities imply the important relation
\bea{relb}
\Del_a^\a\Del_{b\a}W=C_{ac}C_{bd}\,\Delb^{d\ad}\Delb^c_\ad\bar W\ .
\eea

The procedure for reducing $N=2$ superfields and action to $N=1$ form is
well known: one defines $N=1$ superfield components as
\beq
\f\equiv W|\ ,\ \ W_\a\equiv -\Del_{2\a}W|\ ,
\eeq
where the bar denotes setting $\th^{2\a}=\bar\th_2^\ad=0$. One
identifies $\Del_1$ with the $N=1$ covariant spinor derivative, and
rewrites the $N=2$ integration measures in terms of $N=1$ measures
with the replacement
$d^2 \th^2 \rightarrow (\Del_2)^2\equiv\frac12\Del^\a_2\Del_{2\a}$,
{\it etc}.

For a chiral $N=2$ integrand $\FF (W)$ one finds the standard result
\bea{sff}
S_{\FF}&=&\int d^4x \,d^4\th\,\FF (W)= \int d^4x \,d^2\th^1\,\left[
\shalf \FF_{AB}(\Del_2^{ \a}W^A)(\Del_{2\a}W^B) +\FF_A (\Del_2)^2 W^A
\right]|\nonumber\\
&=& \int d^4x \,d^2 \th\, \shalf\FF_{AB}(\f)\,W^{A\a} W^B_\a
+\int d^4x \,d^2\th \,d^2 \bar{\th}\,\FF_A(\f) \,\fb^A\ ,
\eea
where we have used the Bianchi identity (\ref{relb}), dropped the
superscript on $\th^{1\a}$, and, in the last term of the second line,
replaced $\bar{\Del}^2$ by $\int d^2 \bar{\th}$, and we define
$\FF_A\equiv\del\FF/\del\f^A$, {\it etc}. Thus, the holomorphic
contributions to the effective action that we find at the one-loop
level come from $N=2$ chiral integrands (\ref{hol}) (with $\f\to W$).

We turn now to the nonholomorphic contributions.
Consider the reduction to $N=1$ of the $N=2$ superspace integral
\beq
S_\HH=\int d^4\th \,d^4 \bar{\th} \,\, \HH(W, \bar W)\ .
\eeq
Explicitly, we write\footnote{Note that this decomposition of the
$N=2$ measure is real only modulo total derivatives.}
\beq\label{h}
\int d^4\th \,d^4 \bar{\th}\,\, \HH = \int d^2 \th_1 \,d^2 \bar{\th}_1
\, (\Del_2)^2(\Delb^2)^2\,\,\HH |\ .
\eeq
We assume that $\HH$ is gauge invariant up
to total derivatives, which implies:\footnote{There is a similar
issue for $\FF$, which is gauge invariant modulo a quadratic
polynomial with real coefficients \cite{dwetal}.}
\beq\label{hainv}
f^C_{AB}\HH_CW^B + f^C_{AB}\HH_{\bar C}\bar W^B=\eta_A(W)+
\bar\eta_A(\bar W)\ ,
\eeq
where $\eta_A(W)$ are holomorphic functions \cite{bw,hklr}.
We introduce a real function $\mu_A$, the moment map (or Killing
potential), by
\beq\label{mu}
f^C_{AB}\HH_CW^B=\eta_A(W)+i\mu_A(W,\bar W)\ .
\eeq
For the moment, we will set $\eta=0$, and assume $\HH$ is actually gauge
invariant (the calculation with $\eta$ is analogous; see \cite{bw,hklr}).

We expand (\ref{h}) and use the constraints (\ref{cons}) and the
Bianchi identities (\ref{relb}), which imply, in particular,
\bea{rel}
\Del_{2 \a} (\Delb^2)^2 \bar{W}^A| &=&-i\Del_\a^{~\ad}\bar{W}_\ad^A
+f^A_{BC} \fb^B \Del_\a \f^C \ ,\nonumber\\&&\nonumber\\
(\Del_2)^2 \Delb^2_\ad \bar{W}^A| &=& -f^A_{BC} \fb^B \Delb_\ad \fb^C\ ,
\\&&\nonumber\\
(\Del_2)^2(\Delb^2)^2 \bar{W}^A| &=&\shalf \Del^{\a\ad}\Del_{\a\ad}
\fb^A +\shalf f^A_{BC} \fb^B\Del^\a W_\a^C \nonumber \\
&& ~~~~-f^A_{BC} (\Delb^\ad \fb^B)\bar{W}^C_\ad
-f^A_{BC}f^C_{DE} \fb^B\fb^D\f^E\ . \nonumber
\eea
We find
\bea{hh}
S_\HH&=&\int d^4x \,d^2\th \,d^2 \bar{\th}\, \left[
\HH_{\bar{A}\bar{B}CD} \left(\squart W^{D\a}W^C_\a \bar{W}^{B\ad}
\bar{W}^A_{\ad}\right) + \HH_{\bar{A}BC} \left( \shalf W^{C\a}W^B_\a
\Del^2 \f^A \right) \right. \nonumber\\
&&\qquad\qquad\qquad+
\HH_{\bar{A}\bar{B}C} \left( \shalf 
\bar{W}^{B\ad} \bar{W}_{\ad}^A  \Delb^2 \fb^C+i W^{C\a}\bar{W}_\ad^A   \Del_\a^{~\ad}\fb^B
\right)\nonumber\\
&&\qquad\qquad\qquad+\HH_{\bar{A}B}
\left( \Delb^2 \fb^B \Del^2\f^A - f^A_{CD} W^{B\a} \fb^C \Del_\a \f^D +i
W^{B\a} \Del_\a^{~\ad} \bar{W}_\ad^A \right) \nonumber\\
&&\qquad\qquad\qquad+\HH_{\bar{A}\bar{B}}\left( f^A_{CD} \bar{W}^{\ad
B}\fb^C\Delb_{\ad}\fb^D +
\shalf \Del^{\a \ad} \fb^B\Del_{\a \ad}\fb^A \right) \nonumber\\
&&\qquad\qquad\qquad+\HH_{\bar{A}}\left(\shalf \Del^{\a
\ad}\Del_{\a\ad} \fb^A +\shalf f^A_{BC} \fb^B\Del^\a W_\a^C -f^A_{BC}
\bar{W}^{\ad C} \Delb_\ad\fb^B\right.\nonumber\\
&&\qquad\qquad\qquad\qquad\qquad\qquad\qquad\qquad\qquad\left.\left.
-f^A_{BC}f^C_{DE}\fb^B\fb^D\f^E
\right)\right]\ ,
\eea
where $\HH$ and its derivatives are evaluated at
$\th^{2 \a}=\bar\th_2^\ad=0$, and hence are
functions of the $N=1$ chiral superfields $\f$, $\fb$.

Just as for $N=1$ supersymmetric sigma models, this
effective action has a natural K\"ahler geometry: Since the action
doesn't change if we shift $\HH$ by a chiral or antichiral function,
$\HH$ is a K\"ahler potential defined modulo the real
part of a holomorphic function, {\it i.e.}, a K\"ahler transformation.
We introduce the metric, connection, and curvature:
\bea{ggamr}
g_{A\bar B}=\HH_{A\bar B}\ ,\qquad\Gamma^A_{BC}=g^{A\bar
D}\HH_{BC\bar D}\ ,\qquad R_{A\bar BC\bar D}=\HH_{AC\bar B\bar
D}-g_{E\bar F}\Gamma^E_{AC}\Gamma^{\bar F}_{\bar B\bar D}\ .
\eea
Integrating by parts and using (\ref{hainv}) and its derivatives,
we can rewrite the action (\ref{hh}) in a simpler form:
\bea{snice}
S_\HH \!&=& \!\int d^4x \,d^2\th \,d^2 \bar{\th}\,
\left(g_{A\bar
B}\left[-\shalf\Del^{\a\ad}\f^A\Del_{\a\ad}\fb^B
+i\bar W^{B\ad}(\Del^\a{}_\ad W^A_\a
+\Gamma^A_{CD}\Del^\a{}_\ad\f^CW^D_\a)\right.\right.\nonumber\\
&&\qquad\qquad\qquad\qquad~~~
\raisebox{0.0em}[1.em][.9em]{$-(f^A_{CD}\,\bar W^{B\ad}
\f^C\Delb_\ad\fb^D  + f^B_{CD}\,W^{A\a}\fb^C\Del_\a\f^D)$}\\ 
&&\qquad\qquad\qquad\qquad~~~\left.
+(\Del^2\f^B+\shalf\Gamma^{\bar B}_{\bar C\bar D}\bar W^{C\ad}\bar
W^D_\ad) (\Delb^2\fb^A+\shalf\Gamma^A_{EF}W^{E\a}W^F_\a)\right]
\nonumber\\ &&\qquad\qquad\qquad~~\left.
+\squart R_{A\bar BC\bar D}(W^{A\a}W^C_\a\bar W^{B\ad}\bar W^D_\ad)
+i\mu_A(\shalf\nabla^\alpha W^A_\alpha 
+f^A_{BC}\f^B\fb^C) \right)\,\,.  \nonumber
\eea
Because we have rewritten the last term in (\ref{hh}) in terms of the
moment map $\mu$, this expression is valid even when $\eta$
in (\ref{mu}) is nonvanishing.

In $N=1$ language, (\ref{snice}) has contributions to the
effective action of vector and scalar multiplets. None of these
terms modifies component kinetic terms, {\em except\/} the last one,
which is a contribution to the K\"ahler potential. (Clearly, it
vanishes when the superfields $\f$, $\fb$ are restricted to lie in an
abelian subalgebra.) By comparing this last term to the one-loop
contributions we computed above, we can attempt to reconstruct $\HH$.
One can not read off all of $\HH$ from just a knowledge of
this last term, as it is only determined up to an arbitrary function
$f(\f^2,\fb^2)$. However, we know that the $\b$-function and the axial
anomaly come entirely from $\FF$, the holomorphic contribution to the
effective action; this imposes the further constraint that
under an arbitrary complex rescaling of $\f$, $\HH$ changes at most
by a K\"ahler transformation. Up to a K\"ahler transformation, this
implies
\beq\label{hhhom}
\HH=\HH^0+c\left(\ln\frac{\f^2}{\L^2}+g^0(\f)\right)
\left(\ln\frac{\fb^2}{\L^2}+\bar g^0(\fb)\right)\ ,
\eeq
where $\HH^0$, the holomorphic function $g^0$, and its
conjugate $\bar g^0$ are all homogeneous functions and independent
of any scale:
\beq\label{hghom}
\f^A\HH^0_A=\fb^A\HH^0_{\bar A}=\f^Ag^0_A=\fb^A\bar g^0_{\bar A}=0\ ,
\eeq
and $c$ is a constant. The one-loop contributions we have computed
determine $\HH^0$ uniquely. The ambiguity represented by $c$ and $g^0$
is rather minor: the complex manifold with K\"ahler potential $\HH$ and
coordinates $\f$ is a product of a manifold with K\"ahler potential
$\HH^0$ and homogeneous coordinates, and a flat space with a complex
coordinate $Z=\ln(\f^2/\L^2)+g^0(\f)$.

In the case of $SU(2)$, we can determine $\HH^0$ and $g^0$ from
these conditions. In particular, for $SU(2)$, there is no holomorphic
homogeneous gauge invariant function, and so $g^0=0$; thus
the residual ambiguity in $\HH$ is just the constant $c$ in
(\ref{hhhom}).

We consider first the contribution $K_{Q}(\f,\fb)$ in (\ref{hyp}). The
first term is a holomorphic contribution and we consider it no further.
For the second term we introduce the variable
\beq\label{vars}
s= i \frac{|\fb \times \f |}{\fb \cdot \f} = i
\sqrt{\frac{\f^2 \fb^2}{(\f \cdot \fb )^2} -1}\ ,
\eeq
and assume
that $\HH^0 = \HH^0 (s)$ so that $\HH^0_{\bar A} = \HH^0_s s_{\bar A}$ with
\beq
s_{\bar A} =\frac{\del s}{\del\fb^A}=\frac{-1}{s}\left[\frac{\f^2
\fb^A}{(\f\cdot\fb )^2} - \frac{\f^2 \fb^2 \f^A}{(\f \cdot \fb )^3}
\right]
\eeq
Comparing (\ref{hyp}) and (\ref{hh}) we obtain
\beq
\frac{d\HH^0}{ds} = \frac{1}{(8 \pi )^2} \frac{1}{1-s^2}
\ln\frac{1+s}{1-s}
\eeq
which (rewriting the result in terms of $N=2$ superfields) integrates
to
\beq
\HH_Q^{0,fund} (W, \Wbar ) = \frac{1}{(16 \pi )^2} \left[\ln
\frac{\Wbar
\cdot W +i |\Wbar
\times W |}{\Wbar \cdot W -i |\Wbar \times W |} \right]^2
\eeq
For the other contributions in (\ref{ad}) and (\ref{v}) it is more
convenient to introduce
\beq
t= \frac{\fb \cdot \f}{\sqrt{\f^2 \fb^2}}=\frac1{\sqrt{1-s^2}}\ .
\eeq
We find
\beq
\HH_Q^{0,adj}(W, \Wbar ) = \frac{1}{2(4\pi)^2} \int^{T^2}
du\frac{\ln u}{u-1}\ ,\qquad u\equiv t^2
\eeq
and
\beq
\HH^0_V(W, \Wbar ) = \frac{-1}{(8\pi)^2} \left[
 2\left(\ln\frac{T+1}2\right) \left(\ln\frac{T-1}2\right)
+ \int^{T^2}du\frac{\ln u}{u-1}\right]\ ,
\eeq
with
\beq
T = \frac{\Wbar \cdot W}{\sqrt{W^2 \Wbar^2}} .
\eeq

We now discuss our results. We see that the calculation of the
one-loop K\"ahler potential, when combined with the restrictions
imposed by $N=2$ supersymmetry, gives us significant information: we
obtain both the one-loop holomorphic function $\FF$, which determines
the leading terms in the low-energy effective action (terms at most
quadratic in external momenta), as well as (most of) the real function
$\HH$, which contains the next order terms (those at most quartic in
momenta).

The contributions to the one-loop K\"ahler potential
({\it cf.\/} (\ref{hyp}) and (\ref{vec})) can be interpreted in two
ways: When integrating over the loop momentum from zero to some
ultraviolet cut-off, we obtain a finite result because $\f$ acts
as an infrared cut-off for the nonabelian theory, precisely as
described in \cite{s}. This is equivalent to calculating with some
low-energy scale $M$, and integrating out momentum modes \`a la Wilson
between $M$ and the ultraviolet cut-off, provided that $\f  \gg M$.
The physics at momenta below the scale $M$ is described by an {\it
abelian} $N=2$ supersymmetric gauge theory, and is thus encoded in the
holomorphic function $\FF$ as in (\ref{S0}). For an asymptotically
free theory, this is the semiclassical domain where the effective
gauge coupling constant becomes small. In this case, the low-energy
scale $M$ is replaced everywhere by $\f$, and the nonholomorphic
contributions to the K\"ahler potential do not require inverse powers
of $M$ to make them dimensionally correct; instead, extra derivatives
are compensated for by negative powers of $\f$ and
$\fb$, as is clearly shown by the explicit results given above.

Nevertheless, a puzzle seems to remain. When $\f$ is comparable in
size to the low-energy cut-off $M$, we cannot reconcile our results
with $N=2$ supersymmetry. In that case, the K\"ahler potential contains
terms proportional to $[M^2 + \f\fb]\ln [M^2 + \f\fb]$, which cannot
come from a holomorphic function (as required by special geometry)
even when $\f$ and $\fb$ commute. We believe this breaking of $N=2$
supersymmetry arises as follows: It is well known that the effective
action is gauge invariant only if one can shift loop momenta; an
infrared cutoff thus violates gauge invariance. Our quantization
scheme is only $N=1$ supersymmetric, which means that we have $N=2$
supersymmetry only modulo gauge transformations. Therefore, there is
no reason to expect that the effective action with a finite cut-off
$M$, which breaks gauge-invariance, should be $N=2$ supersymmetric. We
stress that $N=2$ supersymmetry is lost only when $\f$ is comparable in
size to the low-momentum cut-off $M$; when $\f$ is either much smaller
or much bigger than $M$, $N=2$ supersymmetry is manifestly realized,
and the K\"ahler potential can be computed from the holomorphic
function $\FF$ and the real function $\HH$.

Our work may be compared to an earlier one \cite{gz} where a one-loop
superspace calculation of the effective potential in $N=1$
supersymmetric theories was presented. Superficially our calculation
is very similar but the emphasis and the contributions we have
considered are different. In ref.\ \cite{gz} supergraphs with
external chiral or Yang-Mills superfields were evaluated, but only
terms with a maximal number of spinor derivatives (and no spacetime
derivatives) outside the loop were kept, thus obtaining a contribution
which is higher order in the auxiliary fields of the chiral multiplets
and/or vector multiplets, and zeroth order in derivatives of the
physical scalars. Here instead, by keeping only terms with a maximal
number of spinor derivatives in the loop, we obtained a contribution
to the K\"ahler potential, which at the component level is quadratic
or lower order in the auxiliary fields and involves space-time
derivatives of the physical fields.

Our results appear to cast doubt on the analysis of \cite{lr}. In
those works, the authors analyzed the holomorphic contributions to the
effective action and concluded that at certain points in moduli space,
some field destabilized because the corresponding terms in the
effective action changed sign. They did not consider possible
modifications due to terms such as those found above. As the
results of \cite{lr} are compatible with arguments based on
nonperturbative formulae for particle masses (BPS bounds),
it would be interesting to explore this issue further.

\begin{flushleft}
{\bf Acknowledgement}
\end{flushleft}

We are happy to thank numerous colleagues
for discussions. We thank R.\ von Unge for reading the manuscript.
BdW thanks the ITP at Stony Brook for hospitality. MG and MR
acknowledge NSF Grants Nos.\ PHY-92-22318 and 93-09888 for partial
support.

\end{document}